# Going from classical to quantum description of bound charged particles: Part 1: Basic Concepts and Assertions

A. L. KHOLMETSKII[1]([*]), T. YARMAN[2] and O. V. MISSEVITCH[3]

[1]*Department of Physics, Belarus State University, Minsk, Belarus*
[2]*Okan University, Istanbul, Turkey & Savronik, Eskisehir, Turkey*
[3]*Institute for Nuclear Problems, Minsk, Belarus*

**Summary**
In this paper we analyze again a transition from the classical to quantum description of bound charged particles, which involves a substantial modification of the structure of their electromagnetic (EM) fields related to the well-known fact that bound micro-particles do not radiate in stationary energy states. We show that a simple exclusion of the radiative component of EM field produced by bound particles leads to a violation of the energy-momentum conservation law, if the non-radiative EM field is left unmodified. In order to restore the energy-momentum conservation, we make a closer look at the interaction of two hypothetical classical charges with the prohibited radiation component of their EM field and bring the appropriate modifications in the structure of their bound EM field and, accordingly, in the Hamilton function of this system. In comparison with the common Hamilton function for the one-body problem, the electric interaction energy is multiplied by the Lorentz factor of orbiting charged particle, and its rest mass $m$ is replaced by an effective rest mass parameter, which includes the interaction EM energy. We introduce, as a novel postulate, these replacements into the Dirac equation for the bound electron and show that the solution of the modified Dirac-Coulomb equation gives the same gross and fine structure of energy levels, as the one furnished by the conventional approach, for hydrogenlike atoms. The correction to spin-spin splitting of 1$S$ state of hydrogen and heavier atoms is much smaller than nuclear structure contribution and can be ignored. However, as discussed in the part 2 of this paper, our approach does induce corrections to the energy levels at the scale of hyperfine interactions, which at once remove a number of long-standing discrepancies between theory and experiment in the atomic physics.

## 1. - Introduction

As a starting point of our analysis, we remind the well-known classical consent that any accelerated charge *must* radiate, and, in particular, both bound (velocity-dependent) and radiating (acceleration-dependent) EM field components do participate in securing the total momentum conservation law for an isolated system "light charge orbiting around heavy charge". In fact, the non-existence of the radiative field component for bound micro-particles (one of Bohr's postulates) historically was the first step to the creation of quantum mechanics. Later, with the development of mathematical apparatus of quantum mechanics and its physical interpretation, it was recognized that the non-applicability of Maxwell equations to a micro-particle looks quite logical, because for such a particle, we even cannot determine the spatial coordinates, velocity and acceleration in the classical meaning. Nonetheless, up to date the usual classical charges with the Maxwellian EM field (representing a composition of non-radiative and free components) are considered as the immediate precursors of bound micro-particles, and a qualitative difference of their EM fields remains unaccounted. A usual way implies that we simply cut off the radiation field component for two interacting charges, but at the same time, no changes are introduced in the Hamilton function and corresponding Hamiltonian of the system. However, the energy-momentum conservation law is not fulfilled for interacting hypothetical classical charges, which produce the non-radiative (bound) EM field only, and thus it seems questionable to leave the Hamilton function of these charges and its quantum counterpart non-modified.

---
[*] E-mails: khol123@yahoo.com

Below we suggest a classical prototype of a system of two bound charges, where the energy-momentum conservation is restored. To accomplish this, we develop a sketch of pure bound field classical electrodynamics (CED), applying a methodological trick as follows. Let us consider classical charged particles, which compose a bound system due to Coulomb interaction, and assume that a motion of these particles can be described in a classical way, but at the same time, a radiation of this system is prohibited. Of course, the Maxwell equations are no longer applicable to this system, and the theory, describing such classical non-radiative charges, is only approximately applicable to real classical phenomena. Nonetheless, the simple rules obtained within such a theory to restore the energy-conservation law in the absence of radiative EM field, occur useful in formulation of an appropriately modified Dirac equation for the case, where the total energy of electron is less than its rest energy (bound state), and in further deducing of a modified Hamiltonians for the description of real quantum particles in a bound state, which do not generate EM radiation (section 2).

As expected, the motional equation for charges in pure bound field CED coincides with the corresponding motional equation for the usual classical charges in the non-relativistic limit, but at the same time, contains some corrections to the accuracy $c^{-2}$, where $c$ being the light velocity in vacuum. Introducing these corrections into the Dirac equation for bound electron and deducing the corresponding Hamiltonian, we find that the solution of the modified Dirac-Coulomb equation gives the same gross and fine structure of energy levels for hydrogenlike atoms, like in the conventional approach (sections 3). In section 4 we analyze quantum two-body problem on the basis of modified Breit equation without external field and show that such an equation yields the same expression for the Dirac-recoil contribution to the atomic energy levels, like in the common approach, to the order $(Z\alpha)^4$, where $\alpha$ is the fine structure constant. In addition, we show that spin-spin hyperfine interaction for hydrogen and heavier atoms remains practically unchanged within the approach we developed (section 5). Finally, section 6 contains a discussion.

## 2. - Pure bound field CED: force law, field equations and motional equation for one-body problem

It is known that the Lagrangian for the system of interacting classical charges represents the sum of three components: matter part, field part and interaction part (see, *e.g.* [1]). It is essential that EM field entering into the Lagrangian includes the bound and free components, and only their sum obeys the Maxwell equations. The same structure of EM field is implied in the conservation laws for the isolated system "source charges plus their EM fields". Hence any attempt to modify the fields without other appropriate changes in the structure of the theory inevitably leads to violation of the conservation laws.

Our nearest goal is to determine the appropriate modifications in the structure of classical electrodynamics, where the EM radiation is prohibited but, at the same time, the energy-momentum conservation law is restored. For our immediate purpose (derivation of new Hamiltonian for quantum non-radiating particles in a physically reasonable way, based on corresponding classical Hamilton function in pure bound field CED), there is no need to develop this pure bound field theory in all details. Usually in the construction of Hamilton function, it is sufficient to determine the momenta of particles and fields, as well as the interaction energy, which in its turn requires that we know the force law applied to the particles. The constraints in question define the nearest tasks to our research: to determine the force law, to postulate the field equations and to derive the energy balance equation in a pure bound field CED. The implementation of these tasks leads us to the solution of the one-body and two-body problems for the bound charge, orbiting without radiation losses around an infinitely heavy charge.

*2'1. Force law in pure bound field CED.* - Due to axiomatic nature of any basic force law, it cannot be deduced in a general way. In this sub-section we consider a particular physical problem, which will help us to postulate the appropriate force law in pure bound CED.

It is well known that the EM field of a classical charged particle is given by the Lienard-Wiechert solution of Maxwell equations, where the field represents the sum of two components: velocity-dependent (bound field) and acceleration-dependent (radiative field). To the accuracy $c^{-2}$, the electric field has the form [1]

$$\boldsymbol{E} \approx \frac{q\boldsymbol{r}}{r^3}\left(1+\frac{\beta^2}{2}-\frac{3(\boldsymbol{\beta}\cdot\boldsymbol{r})^2}{2cr^2}\right)-\frac{q}{2rc^2}\left(\boldsymbol{a}+\frac{(\boldsymbol{a}\cdot\boldsymbol{r})\boldsymbol{r}}{r^2}\right) \tag{1}$$

in the present-time coordinates. Herein $\boldsymbol{v}$ is the velocity, $\boldsymbol{a}$ – the acceleration of a charge and $\boldsymbol{\beta}=\boldsymbol{v}/c$. The physical meaning of the first and second terms in the *rhs* of eq. (1) (bound and free electric field, correspondingly) has been discussed in numerous books and papers (*e.g.*, [1-5]). In particular, it is emphasized that, in general, only the sum of both components represents a solution of Maxwell equations. As a particular demonstration of this general assertion we consider the problem in Fig. 1, which will be useful in further analysis, too.

Within the framework in question, there is an electrically neutral magnetic dipole, which is made of two co-axial homogeneously charged non-conductive rings of almost equal radius $r_0$ and opposite charges $Q$. The negatively charged ring is immovable, whereas the positively charged ring rotates about a common axis of symmetry (the *x*-axis in Fig. 1) at the angular frequency $\omega$. Then one can easily show that, in general, both the velocity dependent component $\boldsymbol{E}_v$ (originated from the first term in *rhs* of eq. (1)) and acceleration dependent component $\boldsymbol{E}_a$ (the second term in *rhs* of eq. (1)) are not vanished separately in an arbitrary point $\{X,0,0\}$, belonging to the *x*-axis. However, the total electric field, representing the sum $\boldsymbol{E}_v+\boldsymbol{E}_a$, is equal to zero.

Now let us consider the same rotating ring in Fig. 1 in pure bound field CED, where the charges do not radiate. This means that for such charges we have to cut off the field terms, which fall off slower than $r^{-2}$: otherwise, the EM energy flux across any sphere with the radius $r$ would reach a finite value at $r\to\infty$, which signifies a loss of the radiation energy. In addition, it is legitimate to assume that the EM field in pure bound field CED coincides with the corresponding velocity-dependent field components in usual CED. With this assumption we come to wipe out the acceleration-dependent term in eq. (1) and obtain the following expressions for the electric $\boldsymbol{E}_b$ and magnetic $\boldsymbol{B}_b$ fields produced by the non-radiative charges:

$$\boldsymbol{E}_b \approx \frac{q\boldsymbol{r}}{r^3}\left(1+\frac{\beta^2}{2}-\frac{3(\boldsymbol{\beta}\cdot\boldsymbol{r})^2}{2cr^2}\right), \quad \boldsymbol{B}_b = \frac{\boldsymbol{v}\times\boldsymbol{E}_b}{c} \tag{2), (3}$$

(hereinafter we supply the quantities obtained within pure bound field CED by the subscript "b").

The difference of $\boldsymbol{E}$ in eq. (1) and $\boldsymbol{E}_b$ in eq. (2) does influence the solution of the problem in Fig. 1, and the *x*-component of the total electric field in the point $\{X,0,0\}$ is no longer vanished, and equal to

$$(E_{total})_x = \beta^2 QX\Big/2\big(X^2+r_0^2\big)^{3/2}. \tag{4}$$

However, this result does contradict the momentum conservation law. It is seen, when we put a charge $q$ into the point $\{X,0,0\}$. The latter experiences the electric force

$$F_x = q\beta^2 QX\Big/2\big(X^2+r_0^2\big)^{3/2}, \tag{5}$$

whereas the reactive force, exerted on the magnetic dipole by the resting charge $q$, is equal to zero, were the standard Lorentz force law assumed. In general, the violation of Newton's third law in EM interactions is not surprising due to a contribution of the momentum of EM field into the total momentum of system. However, the problem in Fig. 1 is stationary, where the EM momentum does not vary with time. Hence the violation of Newton's third law in this case would be unphysical: in particular, if we imagine that the charge and magnetic dipole are fixed on a common platform, then due to the force (5) the platform begins to accelerate along the *x*-axis without an external force.

In order to avoid this situation, one needs to restore the equality of the action and reaction. The simplest way to do this is to modify the Lorentz force law in pure bound field CED and to require that the electric force experienced by a charge in an external EM field, depends on a square of its velocity. In particular, if the velocity-dependent term for this force component is identical to that of eq. (2):

$$\boldsymbol{F}_b^{el(1)} = Q\boldsymbol{E}_b\left(1+\frac{\beta^2}{2}-\frac{3(\boldsymbol{\beta}\cdot\boldsymbol{r})^2}{2cr^2}\right), \tag{6}$$

then the equality of action and reaction is restored for the problem in Fig. 1. Herein $\boldsymbol{E}_b$ is the external electric field at the location of a charge.

If one proceeds from the exact expression for the bound electric field of a moving charge $q$ in the present time coordinates (Heaviside solution [1-3], which we want to keep in pure bound field CED)

$$\boldsymbol{E}_b = \frac{q(1-\beta^2)\boldsymbol{r}}{\left(1-\beta^2\sin^2\vartheta\right)^{3/2}r^3}, \tag{7}$$

a corresponding modification should be introduced into eq. (6), too, which thus reads

$$\boldsymbol{F}_b^{el} = f(\beta^2)q\boldsymbol{E}_b, \quad f(\beta^2) = (1-\beta^2)(1-\beta^2\sin^2\vartheta)^{-3/2}. \tag{8}$$

Herein $\vartheta$ is the angle between the velocity $\boldsymbol{v}$ and the radius-vector $\boldsymbol{r}$ joining the present position of the charge $q$ and the point of observation.

Further, it is natural to assume that the same modification in the force law should be made for the magnetic component, so that we obtain the total force in the form

$$\boldsymbol{F}_b = f(\beta_q^2)\left[q\boldsymbol{E}_b + \frac{q}{c}(\boldsymbol{v}_q\times\boldsymbol{B}_b)\right], \tag{9}$$

where $\boldsymbol{v}_q$ is the velocity of test charge, and $\boldsymbol{\beta}_q=\boldsymbol{v}_q/c$.

In what follows, we will deal with a particular case, where the velocity of the charged particle $\boldsymbol{v}$ is orthogonal to the external electric field $\boldsymbol{E}_b$, so that $\vartheta=\pi/2$. In this case $f(\beta_q^2)=\gamma_q$ ($\gamma_q$ is the Lorentz factor for the test charge $q$), and the electric force component takes the form

$$\boldsymbol{F}_b^\perp = \gamma_q q\boldsymbol{E}_b \text{ (for } \boldsymbol{v}_q\perp\boldsymbol{E}_b\text{).} \tag{10}$$

Hence we can introduce the effective electric field $\boldsymbol{E}_{ef} = \gamma_q\boldsymbol{E}_b$, experienced by a charge with the forbidden EM radiation, moving in the external electric field $\boldsymbol{E}_b$ to be orthogonal to its velocity.

2'2. *Field equations in pure bound field CED.* - It is known that only the sum of bound and radiative EM field components represents the solution of the inhomogeneous Maxwell equations. Our next task is to conjecture their modification in such a way, where the bound component solely would be the solution of the field equations. In this connection we remind that there still exists a particular case, where the non-homogeneous Maxwell equations are implemented for bound EM field along, *i.e.* the case, where the source changes move at constant velocities. In this case the operator $\frac{\partial}{\partial t}=(\boldsymbol{v}\cdot\nabla)$, where $\boldsymbol{v}$ is the constant velocity of source charge. Hence the Maxwell equations take the form as follows:

$$\nabla\cdot\boldsymbol{E}=4\pi\rho, \quad \nabla\cdot\boldsymbol{B}=0, \tag{11a-b}$$

$$\nabla\times\boldsymbol{E}=-\frac{1}{c}(\boldsymbol{v}\cdot\nabla)\boldsymbol{B}, \quad \nabla\times\boldsymbol{B}=\frac{1}{c}(\boldsymbol{v}\cdot\nabla)\boldsymbol{E}+\frac{4\pi\boldsymbol{j}}{c}, \tag{11c-d}$$

where $\rho$ is the charge density, and $\boldsymbol{j}=\rho\boldsymbol{v}$ is the current density. Correspondingly, the inhomogeneous wave equation for the vector potential

$$\Box A = -\frac{4\pi}{c} j, \quad (12)$$

which is valid in common CED [1-3], is replaced by the equation

$$\Delta A - \frac{(v \cdot \nabla)^2}{c^2} A = -\frac{4\pi}{c} j, \quad (13)$$

where $\Box$ is the d'Alembert operator, and $\Delta$ is the Laplacian. Eq. (13) can be also rewritten in the Poisson-like form [5]

$$\Delta A = -\frac{4\pi}{c} j, \quad (14)$$

where the increments $dx$, $dy$, $dz$ in the operator $\Delta = \frac{\partial^2}{\partial x^2} + \frac{\partial^2}{\partial y^2} + \frac{\partial^2}{\partial z^2}$ are related by the Lorentz transformation with the corresponding increments $dx'$, $dy'$, $dz'$ at $dt'= 0$; here the primed quantities belong to the rest frame of source charge (where $j$=0). As known, eqs. (11a-d) and their implication (14) yield the Heaviside solution (7) for EM field of moving charge.

In order to forbid the solutions with radiation component, we adopt that the equations (11) and (14) remain in force in pure bound field CED for an arbitrary velocity of source charges; we see no other way to withdraw the radiative EM field component without violation of the continuity equation and the Lorentz invariance of field equations. From the physical viewpoint, our adoption signifies that in the pure bound field CED framework, the EM field of moving charge keeps the Heaviside form (7) at its arbitrary velocity.

The obtained field equations (11a-d) being complimented by the force law (9), provide a full description of the pure bound field CED.

*2'3. Energy balance equation and EM momentum in pure bound field CED.* - In the standard CED, the EM energy flux density is defined through the Poynting theorem [1-4]

$$\partial u/\partial t + \nabla \cdot S + j \cdot E = 0 \quad (15)$$

in the standard designations.

Now we need to obtain the energy balance equation (analog of Poynting theorem in pure bound field CED), using eqs. (11a-d). We start our analysis with an isolated charged particle, moving with the constant velocity $v$ in the frame of observation. The energy conservation law obviously leads for such a particle:

$$\frac{d}{dt}\int_V u_b dV = 0, \quad (16)$$

where $u_b = \frac{E_b^2 + B_b^2}{8\pi}$ is the EM energy density, and the integration is carried out over the entire free space $V$. Due to the independence of $v$ on spatial coordinates, we further write

$$\frac{du_b}{dt} = \frac{\partial u_b}{\partial t} + (v \cdot \nabla)u_b = \frac{\partial u_b}{\partial t} + \nabla \cdot (v u_b), \quad (17)$$

and combining eqs. (16), (17), we obtain:

$$\frac{\partial u_b}{\partial t} + \nabla \cdot (v u_b) = 0. \quad (18)$$

This equation exactly coincides with the continuity equation in the fluid mechanics and shows that in the present time coordinates EM field rigidly propagates with the source charged particle. The same result follows from the Heaviside expression (7) for EM field of isolated charge, and in a view of its known physical interpretation (e.g., [1-2]), it does not create any problems with respect to causal requirements of classical physics. We also notice that the obtained eq. (18) for an isolated charged particle is equivalent to eq. (15) (see, *e.g.* [6]), with the replacement $u_b \to u$ in pure bound field CED.

Next we analyze the case of two interacting charges $q_1$ and $q_2$, moving with the velocities $v_1$ and $v_2$, which will be sufficient for our purposes. Due to the absence of EM radiation in pure bound field CED, we can choose a large enough spatial volume $V$ enclosing these charges, where the energy flux across the boundary of $V$ due to the bound EM fields becomes negligible. Therefore, the total time derivative of EM energy of this system $E_{EM} = \int_V u_b dV$ must be equal with the reverse sign to the change of kinetic energy of particles:

$$\frac{d}{dt}\int_V u_b dV = -q_1 f(\beta_1^2) v_1 \cdot E_2 - q_2 f(\beta_2^2) v_2 \cdot E_1 = -\int_V \left( f(\beta_1^2) j_1 \cdot E_2 + f(\beta_2^2) j_2 \cdot E_1 \right) dV, \quad (19)$$

where we applied the force law (9), $j_j = q_i \delta(r_i) v_i$ ($i$=1, 2) is the current density for each particle, and

$$u_b = \frac{\left(f(\beta_2^2) E_{b1} + f(\beta_1^2) E_{b2}\right)^2 + \left(f(\beta_2^2) B_{b1} + f(\beta_1^2) B_{b2}\right)^2}{8\pi} =$$

$$= f(\beta_2^2) \frac{E_{b1}^2 + B_{b1}^2}{8\pi} + f(\beta_1^2) \frac{E_{b2}^2 + B_{b2}^2}{8\pi} + f(\beta_1^2) f(\beta_2^2) \frac{E_{b1} \cdot E_{b2} + B_{b1} \cdot B_{b2}}{4\pi}. \quad (20)$$

For fixed arbitrary (large) volume $V$, eq. (19) implies the equality

$$\frac{du_b}{dt} + f(\beta_1^2) j_1 \cdot E_2 + f(\beta_2^2) j_2 \cdot E_1 = 0,$$

The EM energy density $u_b$ in the latter equation represents the function of 7 variables: $u_b = u(t, r_1, r_2)$, where $r_1$, $r_2$ are the present position vectors for each particle. Therefore,

$$\frac{du_b}{dt} = \frac{\partial u_b}{\partial t} + \frac{\partial u_b}{\partial r_1} \cdot \frac{dr_1}{dt} + \frac{\partial u_b}{\partial r_2} \cdot \frac{dr_2}{dt} = \frac{\partial u_b}{\partial t} + \nabla_{r_1}(v_1 u_b) + \nabla_{r_2}(v_2 u_b).$$

Since the flow of EM energy for each particle obeys eq. (18), now we can determine the flow of interactional EM energy, which is the subject of our further interest:

$$\frac{d(u_b)_{in}}{dt} = \frac{\partial(u_b)_{in}}{\partial t} + \nabla_{r_1}(v_1 (u_b)_{in}) + \nabla_{r_2}(v_2 (u_b)_{in}), \quad (21)$$

where, according to eq. (20), $(u_b)_{in} = f(\beta_1^2) f(\beta_2^2)(E_{b1} \cdot E_{b2} + B_{b1} \cdot B_{b2})/4\pi$.

Let us consider the particular case $v_2$=0, which is realized, for example, when the charged particle rotates about another charge with the infinite mass (one-body problem). Hence eq. (21) acquires the form

$$\partial(u_b)_{in}/\partial t + \nabla(v_1 (u_b)_{in}) = 0. \quad (22)$$

This equation determines the interaction EM energy flux density as $(S_b)_{in} = v(u_b)_{in}$. Then, by definition, the interaction EM momentum of the system is

$$P_{EM} = \frac{1}{c^2}\int_V (S_b)_{in} dV = \frac{1}{c^2}\int_V v_1 (u_b)_{in} dV = \gamma_1 \frac{U}{c^2} v_1, \quad (23)$$

where $U$ in the total EM interaction energy, and we have taken into account that $f(\beta_2^2)=1$, and $f(\beta_1^2)=\gamma_1$ for orthogonal $v_1$ and $E_2$.

Now consider the hydrogenlike atom in the classical treatment of pure bound field CED and adopt that the mass of the proton $M$ is finite. In the center-of-mass frame, both particles rotate about their center of mass. Then the interaction EM momentum acquires the form

$$P_{EM} = \gamma_m \gamma_M \frac{U}{c^2}(v_m + v_M). \quad (24)$$

where $v_m$, $v_M$ are the velocities of the electron and proton, correspondingly.

*2'4. Motional equation for charged particles in pure bound field CED (one-body problem).* - Now we are in the position of determining within pure bound field CED a motional equation for a charge *e* with a rest mass *m*, orbiting about a heavy host charge *Ze* with the mass $M\to\infty$, resting in the frame of observation. To handle this problem, we use the law of conservation of the total momentum

$$\boldsymbol{P}_m + \boldsymbol{P}_M + \boldsymbol{P}_{EM} = const, \qquad (25)$$

where $\boldsymbol{P}_m$, $\boldsymbol{P}_M$ are the mechanical momenta of the orbiting and the host particle respectively, and $\boldsymbol{P}_{EM}$ is the interaction EM momentum. Denoting $\boldsymbol{v}$ the velocity of particle *e*, and using eq. (23), we obtain from eq. (25):

$$\frac{d}{dt}\gamma m \boldsymbol{v} + \frac{d}{dt}\frac{\gamma U}{c^2}\boldsymbol{v} = -\frac{d\boldsymbol{P}_M}{dt}, \qquad (26)$$

where

$$\gamma U = -\gamma Z e^2/r. \qquad (27)$$

The *rhs* of eq. (26) represents the force acting on the host particle. For a circular motion of particle *e*, its velocity $\boldsymbol{v}$ is orthogonal to the line joining both particles at any moment of time. Hence according to the Heaviside expression, the electric field $\boldsymbol{E}_b$ of orbiting particle is equal to (see, eq. (7) for $\vartheta = \pi/2$):

$$\boldsymbol{E}_b = \gamma e\boldsymbol{r}/r^3, \qquad (28)$$

at the location of host particle (the origin of co-ordinates), where *r* being the radius of the orbit. Combining eqs. (26)-(28), we obtain

$$\frac{d}{dt}\gamma\left(m + \frac{U}{c^2}\right)\boldsymbol{v} = \gamma\frac{Ze^2\boldsymbol{r}}{r^3}. \qquad (29)$$

The equation of motion (29) written within pure bound field CED differs from the corresponding equation for the one-body problem in usual CED in two points:
- first, the rest mass *m* is replaced by

$$m_b = (m + U/c^2) = mb,$$

where we have introduced the factor

$$b = 1 + U/mc^2;$$

- second, the interaction energy $Ze^2/r$ becomes $\gamma Ze^2/r$.

This induces the introduction of the effective momentum of the particle *e* as

$$\boldsymbol{P}_b = \gamma(m + U/c^2)\boldsymbol{v} = \gamma bm\boldsymbol{v}, \qquad (30)$$

which, in fact, incorporates both the mechanical and EM momenta into the single expression. The same expression for the momentum of bound particle has been introduced by the second author in [7, 8].

The Hamilton function of the system represents the sum of a kinetic energy and a potential energy (27). Hence

$$H = \left(c\sqrt{P_b^2 + m^2 b^2 c^2} - mbc^2\right) - \gamma\frac{Ze^2}{r}, \qquad (31)$$

where $\boldsymbol{P}_b$ is defined by eq. (30).

Now we apply a weak relativistic limit, where the terms of the order $(v/c)^k$ are ignored, if $k>2$. Expanding eq. (31) to the order $(v/c)^2$, one gets:

$$H = \frac{P_b^2}{2mb} - \gamma\frac{Ze^2}{r} - \frac{P_b^4}{8m^3 b^3 c^2}. \qquad (32)$$

Thus, the Hamilton function obtained for the bound charged particles in a weak relativistic limit of pure bound field CED, does differ from the corresponding Hamilton function in usual CED (see, e.g. [1]) by the replacements

$$m \to mb, \quad Ze^2/r \to \gamma Ze^2/r. \qquad (33\text{a-b})$$

In addition, we take into account that the electric field of the nucleus in the one-body problem is equal to $E = -\nabla U/e$, and eq. (33b) implies one more replacement

$$E \to \gamma E. \tag{33c}$$

We again emphasize that the appearance of the replacements (33a-c) is induced by the requirements of energy-momentum conservation in the absence of radiative EM field component. Our principal idea is to extend eqs. (33) to quantum bound systems in the stationary energy states, that guarantees the implementation of energy-momentum conservation for such systems in the absence of EM radiation.

## 3. - Quantum one-body problem

In the quantum domain, the inhomogeneous wave equation (12) is commonly adopted as the counterpart to the classical field equations, where $\hat{A}$ is understood as operator, and the current density $j$ is appropriately re-defined (see, e.g. [9]). However, like in the classical case, this equation is implemented only for the sum of the bound and radiating components of the vector potentials. This means that for the total vector potential $\hat{A} = \hat{A}_b + \hat{A}_f$ (where the subscripts "*b*" and "*f*" denote bound and radiating components of the vector potential, correspondingly), in general[†],

$$\Box \hat{A}_b \neq -\frac{4\pi}{c} j, \quad \Box \hat{A}_f \neq -\frac{4\pi}{c} j.$$

Therefore, eq. (12) *is not applicable* to quantum systems of bound charges with the prohibited radiation, where $\hat{A} = \hat{A}_b$. It seems surprising that, to our recollection, this fact was not commented before, and now we assert, as the novel postulate, that for bound non-radiating charges eq. (12) must be replaced by eq. (14) (where $A$ again represents the operator). Hence the classical system, represented by non-radiating interacting charges and described by the pure bound field CED, is exhibited as an actual classical analog of quantum bound charges, which do not radiate in stationary energy states.

As we have seen above, the elimination of radiative EM field for the system of interacting classical charges implies the appropriate modifications in the Hamiltonian (expressed by eqs. (33)), which aim to secure the total energy-momentum conservation law in the absence of EM radiation. The eqs. (33) give the simple rules for the introduction of the appropriate modifications in quantum mechanical equations for bound non-radiating charges. Of course, such modifications anyway represent a postulate, and only the experiments can validate its correctness.

The replacements similar to (33) can be directly introduced into the Dirac equation in the standard representation (see, *e.g.*, [9, 11]) with a corresponding modification of its solution. However, the approach based on perturbation theory occurs more illustrative for better understanding of physical implications induced by pure bound field theory. In addition, such an approach becomes fruitful for further introduction of corresponding corrections to the atomic energy levels, discussed in the part 2 of the paper.

Thus we start with the Dirac equation (*e.g.* [11]) for the electron bound in the external EM field (when its total energy $E<mc^2$), introducing the quantum counterparts of the replacements (33):

$$\left(i\hbar \frac{\partial}{\partial t} + mb_n c^2 \right)\psi = \left[c\boldsymbol{\alpha}\left(\hat{\boldsymbol{P}}_b - \frac{e}{c} \boldsymbol{A}_b \right) + \beta m b_n c^2 + e\gamma_n U \right]\psi, \tag{34}$$

---

[†] The decomposition of the four-potential into bound and radiating components has been achieved in ref. [10]. We also remind that the homogeneous Maxwell equations can be implemented separately for bound and radiating EM field components, *i.e.* $\Box A_b = 0$, $\Box A_f = 0$.

where $\psi = \begin{pmatrix} \varphi \\ \chi \end{pmatrix}$ is the wave function, $\boldsymbol{\alpha} = \begin{pmatrix} 0 & \sigma \\ \sigma & 0 \end{pmatrix}$, $\beta = \begin{pmatrix} 1 & 0 \\ 0 & -1 \end{pmatrix}$, $\hat{\boldsymbol{P}}_b = -i\hbar\nabla$ is the operator of momentum, $A_b$ is the vector potential, $\sigma$ is the Pauli matrix, and $b_n$, $\gamma_n$ are some coefficients, which represent quantum analogs of corresponding classical factors $b$, $\gamma$. Thus these coefficients are aimed to reflect the non-radiative nature of E M field of bound electron within the total momentum conservation constraint and their divergence from unity has the order of magnitude $(Z\alpha)^2$ and higher. Hence their introduction in the Dirac equation does not affect the gross structure of the energy levels, characterized by the principal quantum number $n$. These coefficients, being constant for any fixed energy level, provide the Lorentz-invariance and other symmetries of relativistic quantum mechanics, when the non-relativistic limit is no longer assumed. We will see below that the coefficients $b_n$, $\gamma_n$ are different in each stationary energy state, and for convenience we supply them by the subscript "$n$".

Further, we imply the presence of electric field only, putting $A=0$. In addition, for a stationary energy state we replace the operator $i\hbar\partial/\partial t$ by the energy $E$ of this state. Then for the functions $\varphi$, $\chi$ we obtain the system of equations as follows:

$$(E - e\gamma_n U)\varphi = c\left(\boldsymbol{\sigma} \cdot \hat{\boldsymbol{P}}_b\right)\chi, \quad (E - e\gamma_n U + 2mb_n c^2)\chi = c\left(\boldsymbol{\sigma} \cdot \hat{\boldsymbol{P}}_b\right)\varphi,$$

with the solution with respect to $\varphi$:

$$(E - e\gamma_n U)\varphi = \frac{c\left(\boldsymbol{\sigma} \cdot \hat{\boldsymbol{P}}_b\right)\varphi}{\left(E - e\gamma_n U + 2mb_n c^2\right)}. \tag{35}$$

In order to derive the equation analogous to the Schrödinger equation in the weak relativistic limit, we introduce the function $\psi$, for which the integral $\int |\psi|^2 dV$ over the entire space does not depend on time. The relationship between the functions $\psi$ and $\varphi$ (see, *e.g.*, [11]) modified due to the substitution (33a) takes the form

$$\psi = \left(1 + \frac{\hat{\boldsymbol{P}}_b^2}{8m^2 b_n^2 c^2}\right)\varphi. \tag{36}$$

Substituting eq. (36) into eq. (35), and implementing the straightforward calculations (see, *e.g.*, [11]), we present the equation for the function $\psi$ in the Schrödinger form $H\psi = E\psi$ with the Hamiltonian

$$H = \frac{P_b^2}{2mb_n} - \gamma_n\frac{Ze^2}{r} - \frac{P_b^4}{8m^3 b_n^3 c^2} - \frac{e\hbar}{2m^2 b_n^2 c^2}\boldsymbol{s}\cdot\left(\gamma_n \boldsymbol{E}_b \times \boldsymbol{P}_b\right) - \frac{e\hbar^2}{8m^2 b_n^2 c^2}(\nabla \cdot \gamma_n \boldsymbol{E}_b), \tag{37}$$

where $\boldsymbol{s}$ is the operator of spin.

The first three terms of this operator represent a quantum counterpart of the Hamilton function (32), the fourth term describes the spin-orbit interaction, while the last term replies to the contact interaction. In this equation we leave out only the terms, which contain the corrections of pure bound field theory at least to the order $(v/c)^{-4}$, as far as the corrections to the higher order $(v/c)^{-6}$ are insignificant at the present measuring accuracy in the atomic physics.

A corresponding Dirac-Coulomb equation acquires the form

$$\left(\frac{P_b^2}{2mb_n} - \gamma_n\frac{Ze^2}{r} - \frac{P_b^4}{8m^3 b_n^3 c^2} - \frac{e\hbar \boldsymbol{s}\cdot\left(\gamma_n \boldsymbol{E}_b \times \boldsymbol{P}_b\right)}{2m^2 b_n^2 c^2} - \frac{e\hbar^2(\nabla\cdot\gamma_n \boldsymbol{E}_b)}{8m^2 b_n^2 c^2}\right)\psi(r,\vartheta,\varphi) = W\psi(r,\vartheta,\varphi), \tag{38}$$

where $\vartheta$, $\varphi$ being the polar and azimuthal angles, correspondingly.

In order to solve eq. (38), we apply the substitution

$$r = r'/b_n\gamma_n, \tag{39}$$

which allows us to rewrite eq. (38) in the convenient form

$$\left[\hat{H}_0(r') + \hat{V}(r')\right]\psi(r',\vartheta,\varphi) = W'\psi(r',\vartheta,\varphi), \quad (40)$$

where $\hat{H}_0(r') = -\dfrac{\hbar^2 \nabla_{r'}^2}{2m} - \dfrac{Ze^2}{r'}$ is the conventional non-relativistic Schrödinger operator, expressed via $r'$-coordinates, whereas

$$\hat{V}(r') = b_n \gamma_n^4 \left( -\frac{\hat{p}_b(r')^4}{8m^3c^2} - \frac{e\hbar\hat{s}\cdot\left(\gamma_n E(r')\times\hat{p}_b(r')\right)}{2m^2c^2} - \frac{e\hbar^2(\nabla_{r'}\cdot\gamma_n E(r'))}{8m^2c^2} \right) \quad (41)$$

is the perturbation, and $W' = W/b_n\gamma_n^2$. Transforming eq. (38) into eq. (41), we used the equalities $\nabla_r = b_b\gamma_n\nabla_{r'}$, $E(r) = b_n^2\gamma_n^2 E(r')$.

Thus, applying the approach of perturbation theory, we first observe that the standard Schrödinger equation

$$\left( -\frac{\hbar^2 \nabla_{r'}^2}{2m} - \frac{Ze^2}{r'} \right)\psi(r',\vartheta,\varphi) = W'\psi(r',\vartheta,\varphi)$$

gives the known solution for the stationary energy levels

$$W'_n = -\frac{mZ^2e^4}{2\hbar^2}\cdot\frac{1}{n^2},$$

as well as the standard Schrödinger–Coulomb wave function $\psi_n(r',\vartheta,\varphi)$ for the hydrogenlike atom. Hence we can apply the known relationships [12]:

$$\overline{v_n^2} = (Z\alpha)^2 c^2/n^2, \quad 1/\overline{r_n} = (Z\alpha/n^2)mc/\hbar,$$

and determine the factors $b_n$, $\gamma_n$ as follows:

$$\gamma_n = \left(1 - \overline{\beta_n^2}\right)^{-1/2} = \left(1 - (Z\alpha)^2/n^2\right)^{-1/2}, \quad b_n = 1 - \frac{Ze^2}{mc^2}\left(1/\overline{r_n}\right) = 1 - (Z\alpha)^2/n^2 \quad (42\text{a-b})$$

for each stationary energy state $n$. Hence we arrive at the equality

$$b_n\gamma_n^2 = 1, \quad (43)$$

at least to the order $(Z\alpha)^2$.

In a view of eq. (43), we get $W' = W$, while the operator of perturbation becomes

$$\hat{V}(r') = \gamma_n^2\left( -\frac{\hat{p}_b(r')^4}{8m^3c^2} - \frac{e\hbar\hat{s}\cdot\left(E(r')\times\hat{p}_b(r')\right)}{2m^2c^2} - \frac{e\hbar^2(\nabla_{r'}\cdot E(r'))}{8m^2c^2} \right). \quad (44)$$

Taking also into account that the perturbation term (44) itself has the order of magnitude $(Z\alpha)^4$, we conclude that our corrections to this term, expressed via the coefficient $\gamma_n$ (42a), appear at least in the order $(Z\alpha)^6$. Hence it is seen that eq. (40) yields the same gross and also fine structure for the atomic energy levels of light hydrogenic atoms, like in the common approach, i.e. the energy of perturbation is expressed by the common equation

$$(\Delta W_b)_n = -\frac{mc^2(Z\alpha)^4}{2n^3}\left(\frac{1}{j+1/2} - \frac{3}{4n}\right) \quad (45)$$

in the standard designations.

Concerning the non-relativistic wave function, we point out that it acquires the standard Schrödinger form in the $r'$-coordinates, related to the laboratory coordinates $r$ by eq. (39). Thus, due to normalization requirement induced by the scaling transformation (39),

$$\psi(r,\vartheta,\varphi) = b_n^{3/2} \gamma_n^{3/2} \psi(r',\vartheta,\varphi). \quad (46)$$

In what follows, we name the theory, which explicitly takes into account the non-radiative nature of EM field of bound charges in stationary energy states, as Pure Bound Field Theory (PBFT), and in the next section we analyze the two-body problem within our approach.

## 4. - Motional equation for bound charged particles (two-body problem) and the structure of energy levels of hydrogenic atoms

In this section we consider the hydrogenlike atom with the finite mass $M$ of the nucleus. Considering first the classical case, we substitute into eq. (25) the EM momentum according to eq. (24). Hence the classical motional equation acquires the form

$$\frac{d}{dt}\gamma_m\left(m + \gamma_M U/c^2\right)v_m = -\frac{d}{dt}\gamma_M\left(M + \gamma_m U/c^2\right)v_M, \quad (47)$$

where the EM interaction energy takes the form

$$U = \frac{1}{4\pi}\int_V (E_m \cdot E_M + B_m \cdot B_M)dV. \quad (48)$$

The equation (47) allows introducing the effective momenta

$$P_{bm} = \gamma_m\left(m + \gamma_M U/c^2\right)v_m, \quad P_{bM} = \gamma_M\left(M + \gamma_m U/c^2\right)v_M, \quad (49\text{a-b})$$

effective masses

$$m_b = m\left(1 + \gamma_M U/mc^2\right) = m b_m, \quad M_b = M\left(1 + \gamma_m U/Mc^2\right) = M b_M \quad (50\text{a-b})$$

and effective interaction energy

$$U_b = \gamma_{mn}\gamma_{Mn}U \quad (50\text{c})$$

of both particles. Herein we have introduced the quantities

$$b_m = \left(1 + \frac{\gamma_M U}{mc^2}\right), \quad b_M = \left(1 + \frac{\gamma_m U}{Mc^2}\right), \quad \gamma_m = \left(1 - v_m^2/c^2\right)^{-1/2}, \quad \gamma_M = \left(1 - v_M^2/c^2\right)^{-1/2}. \quad (51\text{a-d})$$

Introducing the reduced velocity $v_R$ for the two-body problem, eqs. (51c-d) can be also presented in the form

$$\gamma_m = \left(1 - \frac{v_R^2}{c^2}\frac{M^2}{(m+M)^2}\right)^{-1/2}, \quad \gamma_M = \left(1 - \frac{v_R^2}{c^2}\frac{m^2}{(m+M)^2}\right)^{-1/2}. \quad (51\text{e-f})$$

Then the Hamilton function, written in the weak relativistic limit, becomes

$$H = \frac{P_{bm}^2}{2mb_m} + \frac{P_{bM}^2}{2Mb_M} + \frac{P_{bm}^4}{8m^3 b_m^3} + \frac{P_{bM}^4}{8M^3 b_M^3} - \gamma_m\gamma_M\frac{Zq^2}{r}, \quad \text{with } P_{bm} = -P_{bM}.$$

Acting in the same way, like in the previous section, we have to introduce the appropriate modifications in corresponding quantum mechanical equations.

The approach based on the Dirac equation is not directly applicable to the two-particle case, where we should address either to the Bethe-Salpeter equation, or the Breit equation without external field [9], or to their modifications. Though the Breit equation is not fully Lorentz-invariant and represents an approximation, it is the most convenient and illustrative for the analysis of PBFT corrections, resulting due to the replacements (51a-f). Such a re-postulated Breit equation for the Schrödinger-like wave function $\psi(r)$ takes the form

$$\left[\frac{p_b^2}{2mb_{mn}} + \frac{p_b^2}{2Mb_{mn}} - \gamma_{mn}\gamma_{Mn}\frac{Ze^2}{r} - \frac{p_b^4}{8m^3 b_{mn}^3 c^2} - \frac{p_b^2}{8M^3 b_{Mn}^3 c^2} + U_b(p_{bm}, p_{bM}, r)\right]\psi(r) = W\psi(r), \quad (52)$$

where $W$ is the energy, and the term $U_b(p_{bm}, p_{bM}, r)$ is equal to

$$U_b(\boldsymbol{p}_{bm}, \boldsymbol{p}_{bM}, \boldsymbol{r}) = -\frac{\pi Z e^2 \hbar^2}{2c^2}\left(\frac{1}{b_{mn}^2 m^2} + \frac{1}{b_{Mn}^2 M^2}\right)\delta(\boldsymbol{r}) - \frac{Z e^2}{2 b_{mn} b_{Mn} m M r}\left(\boldsymbol{p}_{bm} \cdot \boldsymbol{p}_{bM} + \frac{\boldsymbol{r} \cdot (\boldsymbol{r} \cdot \boldsymbol{p}_{bm})\boldsymbol{p}_{bM}}{r^2}\right) -$$

$$\frac{Z e^2 \hbar \gamma_{mn}\gamma_{Mn}}{4 b_{mn}^2 m^2 c^2 r^3}(\boldsymbol{r}\times\boldsymbol{p}_{bm})\cdot\boldsymbol{\sigma}_m + \frac{Z e^2 \hbar \gamma_{mn}\gamma_{Mn}}{4 b_{Mn}^2 M^2 c^2 r^3}(\boldsymbol{r}\times\boldsymbol{p}_{bM})\cdot\boldsymbol{\sigma}_M - \frac{Z e^2 \hbar \gamma_{mn}\gamma_{Mn}}{2 b_{mn}b_{Mn} m M c^2 r^3}\left((\boldsymbol{r}\times\boldsymbol{p}_{bm})\cdot\boldsymbol{\sigma}_M - (\boldsymbol{r}\times\boldsymbol{p}_{bM})\cdot\boldsymbol{\sigma}_m\right) +$$

$$\frac{Z e^2 \hbar \gamma_{mn}\gamma_{Mn}}{4 b_{mn} b_{Mn} m M c^2}\left[\frac{\boldsymbol{\sigma}_m \cdot \boldsymbol{\sigma}_M}{r^3} - 3\frac{(\boldsymbol{\sigma}_m\cdot\boldsymbol{r})(\boldsymbol{\sigma}_M\cdot\boldsymbol{r})}{r^3} - \frac{8\pi}{3}\boldsymbol{\sigma}_m\cdot\boldsymbol{\sigma}_M\delta(\boldsymbol{r})\right].$$

(53)

We point out that without the introduced PBFT factors $b_{mn}$, $b_{Mn}$, $\gamma_{mn}$ and $\gamma_{Mn}$, eq. (52) acquires its common form [11]. Thus the presence of these factors in eq. (52) determines the PBFT corrections to the Dirac-recoil contribution and spin-spin interval.

In order to solve eq. (52), it is convenient to apply the substitution

$$\boldsymbol{r} = \boldsymbol{r}'/(b_{mn}b_{Mn}\gamma_{mn}\gamma_{Mn}), \qquad (54)$$

which allows us to present the Hamiltonian in eq. (52) as the sum of Schrödinger-like term and perturbation. Indeed, taking into account that $p_b^2 = -\hbar^2 \nabla_r^2 = -b_{mn}^2 b_{Mn}^2 \gamma_{mn}^2 \gamma_{Mn}^2 \hbar^2 \nabla_{r'}^2$, we transform eq. (52) as follows:

$$\left[-\frac{\hbar^2 \nabla_{r'}^2 b_{Mn}}{2m} - \frac{\hbar^2 \nabla_{r'}^2 b_{mn}}{2M} - \frac{Z e^2}{r'} + \frac{1}{b_{mn}b_{Mn}\gamma_{mn}^2\gamma_{Mn}^2}\left(-\frac{p_b^4}{8m^3 b_{mn}^3 c^2} - \frac{p_b^4}{8M^3 b_{Mn}^3 c^2} + U_b(\boldsymbol{p}_{bm},\boldsymbol{p}_{bM},\boldsymbol{r}')\right)\right]\psi(\boldsymbol{r}') = W'\psi(\boldsymbol{r}'), \quad (55)$$

where

$$W' = W/(b_{mn}b_{Mn}\gamma_{mn}^2\gamma_{Mn}^2). \qquad (56)$$

The obtained eq. (55) completed by the expressions (53), (54) and (56), represents the basic equation for the quantum two-body problem within the framework of PBFT. Here one should recall that eq. (55) itself, like the source Breit equation, is semi-relativistic, and it is valid to the order $(Z\alpha)^4$. At the same time, the factors $b_{mn}$, $b_{Mn}$, $\gamma_{mn}$ and $\gamma_{Mn}$, being explicitly determined to the orders $(Z\alpha)^2$ and $(Z\alpha)^4$ (see below), allow us to analyze the specific PBFT corrections to the order $(Z\alpha)^6$, which correspond to the scale of hyperfine interactions. The determination of these corrections is the next goal of our analysis, but, first of all, let us show that eq. (55) yields the same gross and fine structure of the atomic energy levels, as the one furnished by the common approach.

In the zeroth approximation, when the terms of order $(v/c)^2$ and higher are ignored, we get from eq. (55) the Schrödinger equation expressed in $\boldsymbol{r}'$-coordinates:

$$\left(-\frac{\hbar^2 \nabla_{r'}^2}{2m_R} - \frac{Z e^2}{r'}\right)\psi(\boldsymbol{r}') = W\psi(\boldsymbol{r}'),$$

where $m_R = mM/(m+M)$ is the reduced mass. Hence we obtain the well-known solution

$$W_{0n} = -\frac{m_R c^2 (Z\alpha)^2}{2 n^2},$$

along with the common Schrödinger wave function expressed via $\boldsymbol{r}'$-coordinates. This result allows us to obtain the coefficients $b_{mn}$, $b_{Mn}$, $\gamma_{mn}$, $\gamma_{Mn}$ at least to the order $(Z\alpha)^2$, based on their respective classical limits (51a-f), taking into account the known relationships (*e.g.* [12]):

$$\frac{\overline{U}}{mc^2} = -\frac{Z e^2}{\overline{r}\, mc^2} = -\frac{Z e^2}{\overline{r}\, m_R c^2}\frac{M}{M+m} = -\frac{(Z\alpha)^2}{n^2}\frac{M}{M+m}, \qquad (57a)$$

$$\frac{\overline{U}}{Mc^2} = -\frac{Ze^2}{\overline{r}\,Mc^2} = -\frac{Ze^2}{\overline{r}\,m_R c^2}\frac{m}{M+m} = -\frac{(Z\alpha)^2}{n^2}\frac{m}{M+m}, \qquad (57b)$$

$$\frac{\overline{v_m^2}}{c^2} = \frac{\overline{v_R^2}}{c^2}\frac{M^2}{(M+m)^2} = \frac{(Z\alpha)^2}{n^2}\frac{M^2}{(M+m)^2}, \quad \frac{\overline{v_M^2}}{c^2} = \frac{\overline{v_R^2}}{c^2}\frac{m^2}{(M+m)^2} = \frac{(Z\alpha)^2}{n^2}\frac{m^2}{(M+m)^2}. \qquad (57\text{c-d})$$

Hence, via the comparison of eqs. (57a-d) with eqs. (51a-b, e-f), we obtain the factors $b_{mn}$, $b_{Mn}$, $\gamma_{mn}$, $\gamma_{Mn}$ to the accuracy $(Z\alpha)^2$ as follows:

$$b_{mn} = \left(1 - \frac{(Z\alpha)^2}{n^2}\frac{M}{M+m}\right), \quad b_{Mn} = \left(1 - \frac{(Z\alpha)^2}{n^2}\frac{m}{M+m}\right), \qquad (58\text{a-b})$$

$$\gamma_{mn} = \left[1 - \frac{(Z\alpha)^2}{n^2}\frac{M^2}{(m+M)^2}\right]^{-1/2}, \quad \gamma_{Mn} = \left[1 - \frac{(Z\alpha)^2}{n^2}\frac{m^2}{(m+M)^2}\right]^{-1/2}. \qquad (58\text{c-d})$$

Further on, using eqs. (58a-d), we derive the product

$$b_{mn} b_{Mn} \gamma_{mn}^2 \gamma_{Mn}^2 = 1 - \frac{(Z\alpha)^2}{n^2}\frac{2mM}{(M+m)^2} \qquad (59)$$

to the accuracy of calculations $(Z\alpha)^2$.

Applying equations (58), (59), as well as the equality [11]

$$\overline{p^2} = 2m_R\left(\overline{W_{on} + \frac{Ze^2}{r}}\right) = m_R^2 c^2 (Z\alpha)^2/n^2, \qquad (60)$$

we find that to the accuracy of calculations $(Z\alpha)^4$,

$$\frac{p_b^2(r')b_{Mn}}{2m} + \frac{p_b^2(r')b_{mn}}{2M} - \frac{W}{b_{mn}b_{Mn}\gamma_{mn}^2\gamma_{Mn}^2} = \frac{p^2(r')}{2m} - W. \qquad (61)$$

Substituting this equality into eq. (55), and ignoring the PBFT factors $b_{mn}$, $b_{Mn}$, $\gamma_{mn}$, $\gamma_{Mn}$ in the terms of the order $(Z\alpha)^4$, we obtain:

$$\left[-\frac{\hbar^2 \nabla_{r'}^2}{2m_R} - \frac{Ze^2}{r'} + \left(-\frac{p^4}{8m^3 c^2} - \frac{p^4}{8M^3 c^2} + U(\boldsymbol{p}_m, \boldsymbol{p}_M, \boldsymbol{r}')\right)\right]\psi(\boldsymbol{r}') = W\psi(\boldsymbol{r}'), (62)$$

where the term $U(\boldsymbol{p}_m, \boldsymbol{p}_M, \boldsymbol{r}')$ differs from the term $U_b(\boldsymbol{p}_m, \boldsymbol{p}_M, \boldsymbol{r}')$ in eq. (55) by the omission of PBFT factors $b_{mn}$, $b_{Mn}$, $\gamma_{mn}$, $\gamma_{Mn}$.

Excluding further the spin-spin interaction in the expression for $U(\boldsymbol{p}_m, \boldsymbol{p}_M, \boldsymbol{r}')$ (last term in the *rhs* of eq. (53)), we arrive at the common solution for the Dirac-Recoil (DR) contribution to the energy levels, written to the order $(Z\alpha)^4$ [11]:

$$\left(W_b^{DR}\right)_{nlj} = m_R c^2 \left\{[f(n,j) - 1] - \frac{m_R}{2(m+M)}[f(n,j) - 1]^2\right\}, \qquad (63)$$

where $f(n,j) \approx 1 - \frac{(Z\alpha)^2}{2n^2} - \frac{(Z\alpha)^4}{2n^3}\left(\frac{1}{j+1/2} - \frac{3}{4n}\right)$, and $j$ is the quantum number of total angular momentum ($j=l+s$, $l$ is the angular momentum, and $s$ the electron's spin).

Thus the corrections of PBFT to the Dirac-recoil contribution may emerge at least in the order $(Z\alpha)^6$, which corresponds to hyperfine interactions, and which will be determined in the part 2 of this paper.

## 5. - Hyperfine splitting of energy levels due to spin-spin interaction in hydrogen and heavier atoms

Now we show that the corrections to the term of spin-spin interaction $(W_{HFS})_b$ within PBFT are well below the present experimental uncertainty.

First we analyze the contribution of spin-spin interaction into the Breit potential, which in PBFT has the form

$$(V_b(r))_{s-s} = \frac{1}{b_{mn}b_{Mn}\gamma_{mn}^2\gamma_{Mn}^2} \frac{Ze^2 h^2 \gamma_{mn}\gamma_{Mn}}{4mb_{mn}Mb_{Mn}c^2}\left(\frac{\boldsymbol{\sigma}_m \cdot \boldsymbol{\sigma}_M}{r^3} - 3\frac{(\boldsymbol{\sigma}_m \cdot \boldsymbol{r})(\boldsymbol{\sigma}_M \cdot \boldsymbol{r})}{r^5} - \frac{8\pi}{3}\delta(r)\right) \quad (64)$$

(the last term of eq. (53)). Being expressed via $r'$-coordinates, this operator reads:

$$(V_b(r'))_{s-s} = b_{mn}b_{Mn}\gamma_{mn}^2\gamma_{Mn}^2 \frac{e^2 h^2}{4mMc^2}\left(\frac{\boldsymbol{\sigma}_m \cdot \boldsymbol{\sigma}_M}{r'^3} - 3\frac{(\boldsymbol{\sigma}_m \cdot \boldsymbol{r'})(\boldsymbol{\sigma}_M \cdot \boldsymbol{r'})}{r'^5} - \frac{8\pi}{3}\delta(r')\right), \quad (65)$$

where we have used eq. (54) and the relationship $\delta(r) = b_{mn}^3 b_{Mn}^3 \gamma_{mn}^3 \gamma_{Mn}^3 \delta(r')$. Designating

$$(V(r'))_{s-s} = \frac{e^2 h^2}{4mMc^2}\left(\frac{\boldsymbol{\sigma}_m \cdot \boldsymbol{\sigma}_M}{r'^3} - 3\frac{(\boldsymbol{\sigma}_m \cdot \boldsymbol{r'})(\boldsymbol{\sigma}_M \cdot \boldsymbol{r'})}{r'^5} - \frac{8\pi}{3}\delta(r')\right)$$

(the common Hamiltonian of spin-spin interaction expressed via $r'$-coordinates), and using eq. (59), we obtain

$$(V_b(r'))_{s-s} = \left(1 - \frac{(Z\alpha)^2}{n^2}\frac{2mM}{(M+m)^2}\right)(V(r'))_{s-s}. \quad (66a)$$

This relationship is also valid for the energy of spin-spin interaction, obtained via the averaging of operators $(V_b(r'))_{s-s}$ and $(V(r'))_{s-s}$ with the wave function $\psi(r')$:

$$(W_b)_{s-s} = \left(1 - \frac{(Z\alpha)^2}{n^2}\frac{2mM}{(M+m)^2}\right)W_{s-s}. \quad (66b)$$

Thus the term

$$\delta(W_b)_{s-s} = -\frac{(Z\alpha)^2}{n^2}\frac{2mM}{(M+m)^2}W_{s-s} \quad (67)$$

determines the PBFT correction to hyperfine splitting.

For the 1S state of hydrogen, the term $\alpha^2 \frac{2mM}{(M+m)^2}W_{s-s}$ of eq. (67) itself is less than 100 Hz (where we have used the measured value $W_{s-s}$=1 420 405.751 768(1) kHz [13]) and is many times smaller than the nuclear-structure corrections to the 1S hyperfine splitting, which vary from tens to hundreds kHz [14-16]. Thus for hydrogen the PBFT correction to spin-spin splitting occurs negligible, and we put

$$(W_b)_{s-s}^H = W_{s-s}^H. \quad (68)$$

within the range of the present uncertainty in calculation of $W_{s-s}^H$.

For heavier atoms, the PBFT correction (67) becomes smaller, whereas the nuclear structure effects are larger. Hence the correction (67) can be well ignored for all such atoms, too.

The case of leptonic atoms occurs more complicated, and it will be analyzed separately in the part 2 of the present paper.

## 6. - Discussion

Thus, the qualitative difference between classical and quantum systems of bound charged particles with respect to their ability to emit electromagnetic radiation, makes a transition from the classical to quantum description of such systems more complicated than it was originally conceived. Our principal assertion is that the Hamilton function written for interacting classical charges should be modified before constructing a corresponding Hamiltonian for wave-like bound particles. This assertion is closely related to the non-applicability of the non-homogeneous wave equation (12) to quantum bound systems and to the requirement of energy-momentum conservation for such systems, when their EM radiation is forbidden. Hence a logically non-contradictory transition from classical to quantum description of bound charges should proceed from the classical Hamilton function to its quantum counterpart, where both bear a common fundamental structure of EM field, consisting of only the bound field component (eq. (14)). In order to describe such a transition at the mathematical level, we have developed a pure bound field CED, where the motion of charges is described in a classical way, but their EM radiation is forbidden. The Hamilton function written within such a pure bound field CED differs from the conventional Hamilton function in standard CED in two points: the rest masses $m$ and $M$ of two interacting particles are replaced by the effective rest mass parameters $m+\gamma_M U/c^2$, $M+\gamma_m U/c^2$, correspondingly, and the interaction energy is replaced by $\gamma_m \gamma_M U$. The appearance of the Lorentz factors $\gamma_m$ and $\gamma_M$ in these expressions reflects the dynamics of particles in pure bound field CED under the energy-momentum conservation constraint.

The introduction of the masses $m_b$ and $M_b$ into the corresponding Hamilton operator with the accompanied replacement $U \rightarrow \gamma_m \gamma_M U$, leads us to the modified Dirac-Coulomb equation for the quantum one-body problem, and to the modified Breit equation without external field for the quantum two-body problem, giving the results as follows:
- the gross structure of the hydrogenlike atoms is not influenced by PBFT, because the PBFT corrections to the equations of atomic physics have the order of magnitude $(Z\alpha)^2$ and they cannot affect the gross structure by definition;
- the Dirac-recoil contribution to the energy levels also coincides with the corresponding expression in the standard theory, at least to the order of magnitude $(Z\alpha)^4$, eq. (63);
- the PBFT correction to the hyperfine spin-spin interaction for the hydrogen and heavier atoms (67) occurs much less that the present calculation uncertainty and thus can be ignored.

This is an important step of validation of PBFT, giving the required coincidence of the fine structure and spin-spin interaction with well-proved experimental data.

We emphasize that the results listed above have been obtained with the scaling transformation (54), which can be interpreted as the increase of form-factors for hydrogenic atoms by $b_{mn} b_{Mn} \gamma_{mn} \gamma_{Mn}$ times in comparison with the commonly adopted value. In the classical analogy, this effect is explained by the reduction of the effective rest masses of orbiting particles by $b_m$ and $b_M$ times, correspondingly (see eq. (50)), that causes an increase of the radius of their orbits. In the part 2 we will further explore the physical meaning of the transformation (54) and its implications. At the same, one should take into account that the product $b_{mn} b_{Mn} \gamma_{mn} \gamma_{Mn}$ differs from unity in the order $(Z\alpha)^2$ and higher, which is well below of the present experimental uncertainty in the measurement of form-factors for light hydrogenic atoms.

Further, it would be fair to bring up that this work is initiated based on an idea of the second author that the rest mass of any object bound to a given field should be decreased as much as the mass equivalent of the "static binding energy" coming into play (and this, for classical particles, already at rest) [7, 8]. However, a detailed discussion of this idea falls outside the scope of the present contribution.

As a final remark, it is worth to emphasize that the purely bound field CED was conceived along with a single purpose: using standard quantization scheme, to take explicitly into account a modification of EM field of charged particle in bound quantum state. Thus, such a bound field theory is not a substitution for the conventional CED and hence does not imply any change of the recognized limits of CED applicability. Rather our goal is to introduce into the Hamiltonian the appropriate corrections, which reflect a non-radiative nature of EM fields of bound wave-like particles.

In the part 2 of the paper we will show that PBFT evokes significant corrections to the energy levels computed within QED at the range of hyperfine contributions to the atomic energy levels. As an important outcome, we will remove the long-standing discrepancies between theory and experiment in physics of light hydrogenic atoms: 1*S*-2*S* interval for positronium, hyperfine spin-spin splitting of 1*S* level in positronium, classic Lamb shift and the ground state Lamb shift in the hydrogen. In particular, we will show that the proton charge radius derived within the PBFT framework, is in a perfect agreement with the result of its latest (and the most precise) measurement via the 2S-2P Lamb shift in muonic hydrogen [17].

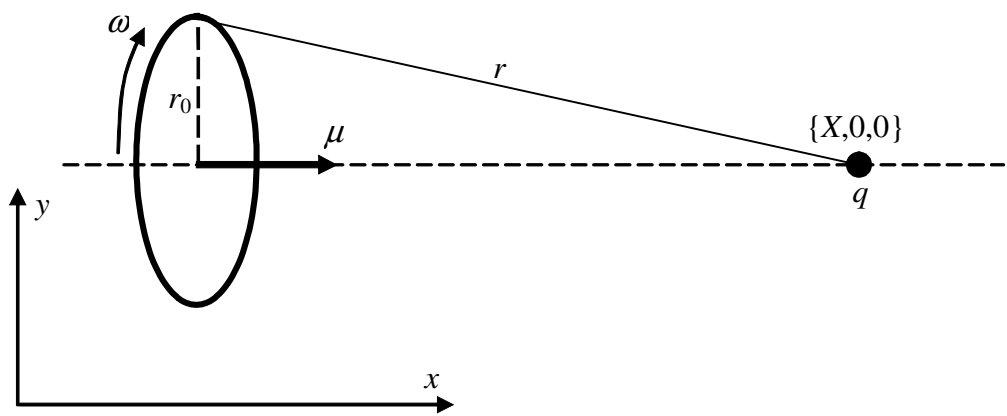

Fig. 1. Interaction of the magnetic dipole **μ** and the charge *q*, located on the axis of the dipole.